\documentclass[11pt]{article}
\usepackage{moriond,epsfig}
\usepackage[latin1]{inputenc}
\usepackage[OT1]{fontenc}

\def\lsim{\mathrel{\hbox{\rlap{\hbox{\lower4pt\hbox{$\sim$}}}\hbox{$<$}}}}
\def\gsim{\;\raise0.3ex\hbox{$>$\kern-0.75em\raise-1.1ex\hbox{$\sim$}}\;}

\def\be{\begin{equation}}
\def\ee{\end{equation}}
\def\bea{\begin{eqnarray}}
\def\eea{\end{eqnarray}}

\begin{document}
\vspace*{3.2cm}

\title{%
\vskip-6pt \hfill {\rm\normalsize CERN-PH-TH/2008-101} \\
\vskip-12pt~\\
EXPERIMENTAL RELEVANCE OF LOW REHEATING TEMPERATURE COSMOLOGIES~\footnote{
Talk given at the 43rd Rencontres de Moriond- Cosmology, La Thuile, Italy, March 15 - 22, 2008.}}

\author{GRACIELA B. GELMINI}

\address{Department of Physics and Astronomy, UCLA,\\
475 Portola Plaza, Los Angeles, CA 90095, USA;\\
CERN PH-TH, CH-1211, Geneva 23, Switzerland}

\maketitle\abstracts{
Standard simple  assumptions are usually made about the pre-Big Bang
Nucleosynthesis epoch, from which we do not have observations.
Modifying these assumptions, the predicted density of relic particles such as
neutralinos and sterile neutrinos can be very different from that in the standard case.
  For example, neutralinos could have the dark matter density
in (almost) any supersymmetric model, and sterile neutrinos with mixings large
enough to be soon detected  in  neutrino experiments would become
cosmologically acceptable. These possibilities  are important 
in view of what the LHC, and neutrino experiments
could soon find.  }

The argument showing that WIMPs (weakly interacting massive particles) are good dark matter  (DM) candidates is old. The density per comoving volume of non relativistic  particles in equilibrium in the early Universe decreases exponentially with decreasing temperature, due to the Bolzmann factor, until the reactions which change the particle number become ineffective.  
 At this point, when  the annihilation rate  becomes smaller than the Hubble expansion rate, 
  the WIMP number per comoving volume becomes constant. This  moment of chemical decoupling or freeze-out happens later, i.e. for smaller WIMP densities, for larger annihilation cross sections $\sigma$. If there is no subsequent change of entropy in matter plus radiation, the present relic density is $\Omega_{\rm std} h^2 \simeq 10^{-10} {\rm ~GeV^{-2}}/ {\left< \sigma v \right> } $, which for weak cross sections gives the right order of magnitude of the DM density (and a temperature  $T_{f.o.} \simeq m/20$ at freeze-out for a WIMP of mass $m$). 

 This is a ballpark argument. When actually applied to particle models, the requirement that the WIMP candidate of the model must have the measured DM density is very constraining. In many supersymmetric models, in which the WIMP candidate is usually a neutralino, this ``DM constraint" 
is very effective in restricting the parameter space of models. In minimal supergravity models (mSUGRA) for instance, the neutralino typically has a  small annihilation rate in the early Universe, thus its relic density  tends to be larger than observed. The  ``DM constraint"  is found to  be satisfied only along  four very  narrow regions in  the fermionic and scalar mass parameter space $m_{1/2}$, $m_0$ (see e.g. Ref.~1): the ``bulk" (with a light neutralino and tight
accelerator constraints), the ``coannihilation region" (where coannihilations with a stau suppress the
relic density), the ``funnel region" (where  resonance effects enhance the  neutralino-neutralino annihilation rate) and the ``focus point region" (where the neutralino acquires a non-negligible higgsino fraction).   Most of the ``benchmark points", special models chosen to study in detail in preparation for the  LHC  and the next possible collider 
(such as A' to L', Snowmass Points and Slopes or SPS 1a',1b, 2, 3 ,4, 5,  Liner Collider Cosmo points or LCC 1,2,3,4) lie on those very narrow bands (which become more fine-tuned for large $m_{1/2}$ and $m_0$ values)~\cite{benchmarks}. Neutralinos are underabundant (account for a fraction of the DM) also in narrow regions adjacent to these just mentioned, but  in most of the parameter space neutralinos are overabundant and the corresponding models are rejected.
Is it  correct to reject all these supersymmetric  models? 

 The issue is that the narrow bands just mentioned  depend not only on the particle model to be tested in collider experiments, but on the history of the Universe before Big Bang Nucleosynthesis (BBN), an epoch from which we have no data. BBN is the earliest episode (200 s  after the Bang,  $T\simeq 0.8$ MeV) from which we have a trace, the abundance of light elements
D, $^4$He and $^7$Li. The next observable is the Cosmic Microwave Background radiation (produced 3.8 $~10^4$ yr after the Bang, at $T\simeq$ eV) and the next is the Large Scale Structure of the Universe. WIMP's have their number fixed at $T_{f.o.} \simeq m/20$, thus WIMPs with $m \gsim 100$ MeV would be the earliest remnants and, if discovered,  they would  for the first time give  information on the pre-BBN phase of the Universe.

As things stand now, to compute the WIMP relic density we must make assumptions about the pre-BBN epoch. The standard computation of the relic density relies on the standard
assumptions that the entropy of matter and radiation is conserved,
that WIMPs are produced thermally, i.e. via interactions with the particles in the plasma,  and were in kinetic and chemical equilibrium before they decoupled at $T_{f.o.}$. These are just assumptions, which do not hold in many cosmological models. These include models with moduli decay, Q-ball decay and thermal inflation \cite{Moroi-etc}, in which there is a late episode of entropy production or inflation and  non-thermal production of the WIMPs in particle decays is possible. It is enough that the
highest temperature of the radiation dominated period in which BBN happens, the so called
reheating temperature $T_{RH}$, is larger than 4 MeV~\cite{hannestad} for BBN and all the subsequent  history of the Universe to proceed as usual.
In non-standard cosmological models the WIMP relic abundance may be decreased  or increased with respect to the standard abundance. The density may be decreased by reducing the rate of  thermal production (through a low $T_{RH} < T_{f.o.}$) or by producing radiation after freeze-out (entropy dilution). The density may be increased by creating WIMPs from particle (or extended objects) decay (non-thermal production) or by increasing the expansion rate of the Universe at freeze-out (e.g. in kinetion domination models~\cite{kinetion}). 

In models in which a  scalar field $\phi$ dominates the energy density of the Universe and decays late  producing a plasma with a low temperature $T_{RH}$,  the WIMP density depends only on two additional parameters besides the usual ones~\cite{gg}. These parameters depend on the completion of the  models to higher energy scales that those that will be tested in colliders. They are $T_{RH}$ and the ratio  of the average  number $b$ of WIMPs produced per $\phi$ decay and the $\phi$ mass $m_\phi$, $\eta= b 100{\rm TeV} / m_\phi$.  $\phi$ could be one of the moduli fields pervasive in string or plain supersymmetric models or an inflaton field.
 \begin{figure}
\hspace{1.0cm}
\includegraphics[width=0.90\textwidth]{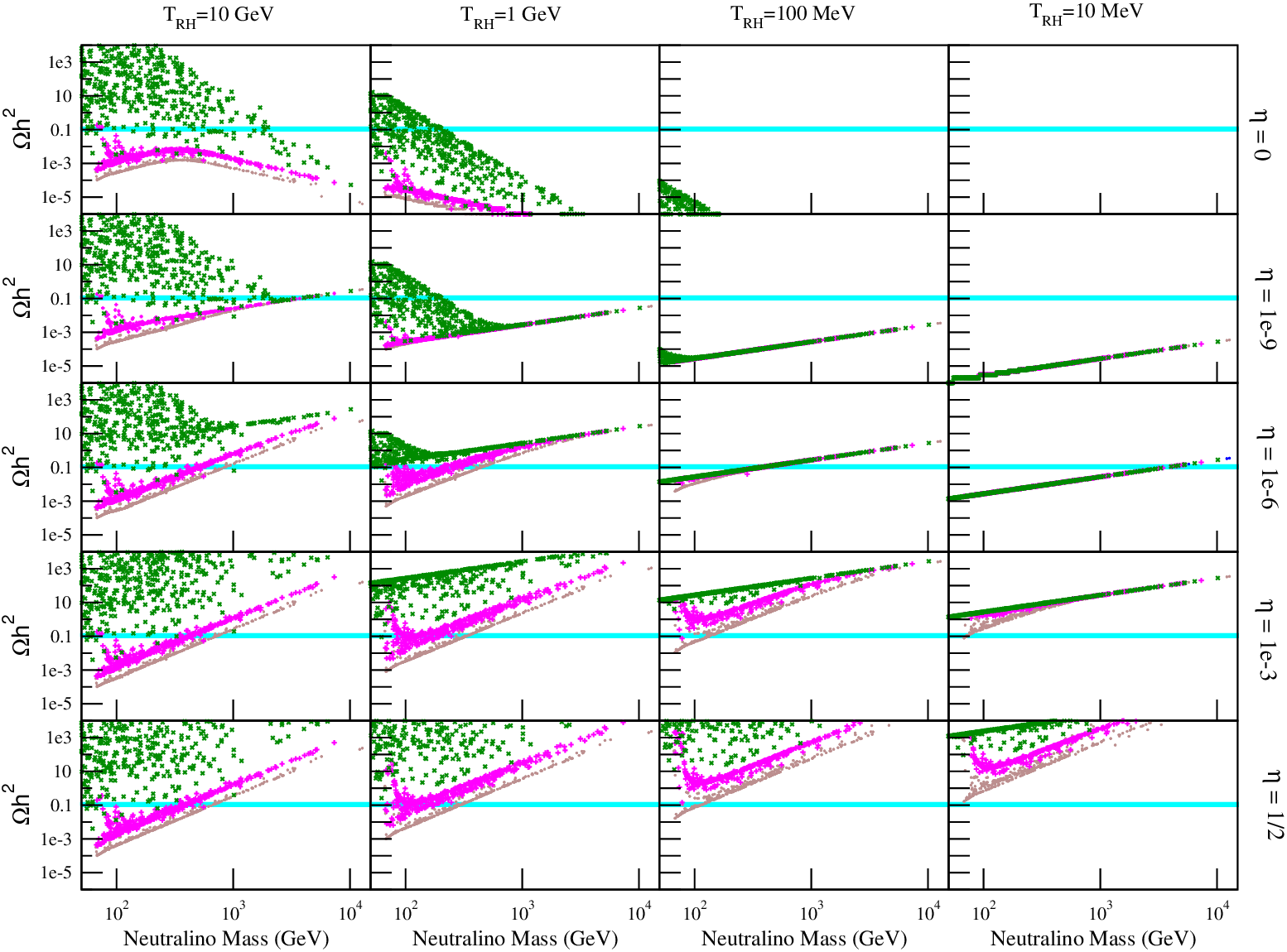}
\vspace{-10pt}
\caption{Neutralino relic density $\Omega h^2$ vs mass for different values of $T_{RH}$ and $\eta$ for 1700 different MSSMs, from Ref.~7 (color indicates composition: green for bino-like, pink for higgsino-like and brown for wino-like neutralinos).}
\vspace{-10pt}
\end{figure}
 Fig.1 shows the relic density as function of the mass of neutralinos in 1,700 different Minimal Supersymmetric Standard Models (MSSMs) characterized by nine parameters defined at the electroweak scale~\cite{ggsy}, each shown as one point in each of the panels. The upper-left panel  (corresponding to high $T_{RH}$ and $\eta=0$) shows the standard density, which can be either higher than the DM range, shown in cyan, or lower or just right. The figure shows that all points can be brought to cross the DM cyan line with suited combinations (in general not unique) of $T_{RH}$ and $\eta$. This means that neutralinos can have the DM density in (almost) all supersymmetric models, provided the right values of $T_{RH}$ and $\eta$ can be obtained (the exception being severely overabundant or underabundant very light neutralinos~\cite{ggsy-2}, rarely encountered in supersymmetric models). This has important implications not only for colliders but for direct and indirect DM searches as well.

We see in Fig. 2.a the standard relic density region as function of the mass of $10^5$ MSSMs defined by 10 parameters at the electroweak scale~\cite{ggsy-2}. Here the bino mass parameter $M_1$ was allowed be much smaller than the other two gaugino mass parameters, as low as 100 MeV, which does not contradict any experimental bound (in models in which the three gauginos masses are not required to coincide at a large energy scale). All the models above the black strip showing the right DM abundance are rejected in the standard cosmology, because the neutralinos they predict are overabundant. Only the models in the black and red regions (with the right DM density or less) are usually assumed to be viable. Their halo fraction times spin-independent proton-neutralino cross section  $f \sigma_{SI}$ (on which the interaction rate  in direct detection experiments depends) falls in the red-black region shown in Fig.2.b, which extends from 30 GeV to 2 TeV in neutralino mass. Accepting all the models shown in Fig.2.a whose density can be brought to coincide with the DM density, and assuming that they have the DM density, the region of viable supersymmetric models to be searched for in direct detection experiments changes to the blue region of Fig.2.b, which extends from  under 1 GeV to  10 TeV in neutralino mass.
Shown in Fig.2.b are also the present experimental limits (black solid line) and future discovery limits of several direct DM search experiments (see Ref.~8 for details). 
\begin{figure}
\includegraphics[width=0.49\textwidth]{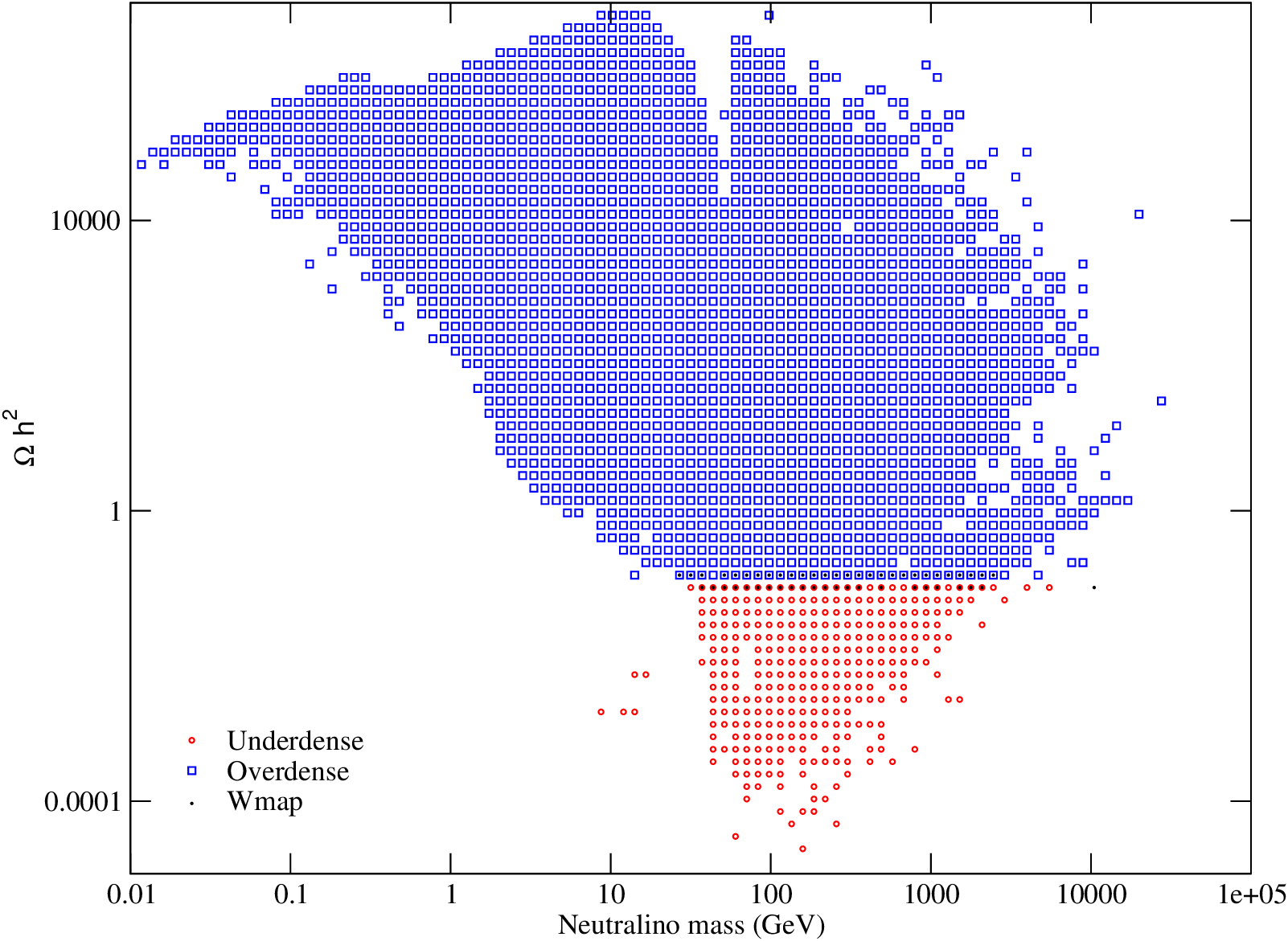}
\includegraphics[width=0.49\textwidth]{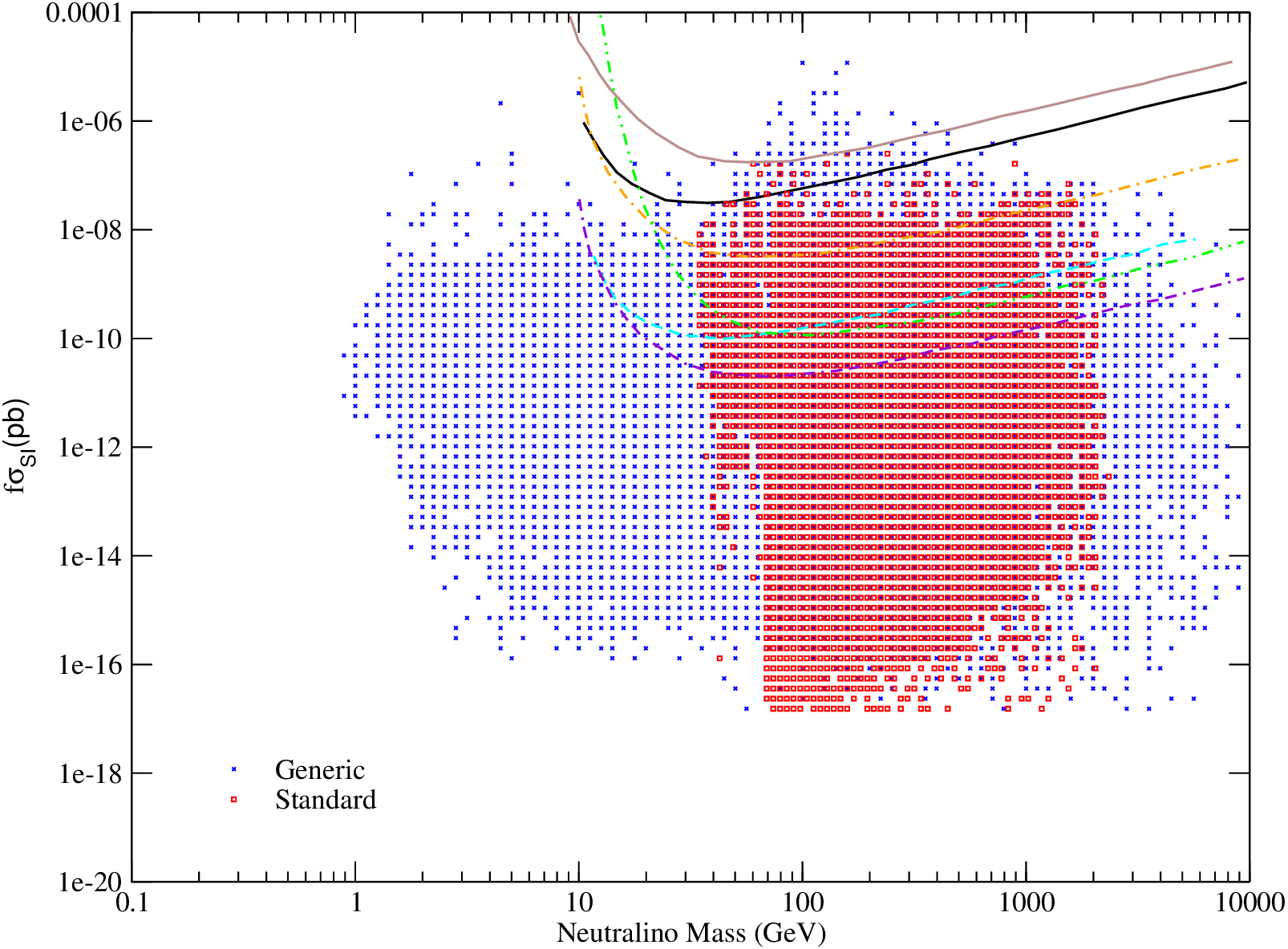}
\vspace{-10pt}
\caption{a.(left) Standard neutralino density $\Omega h^2$ and b.(right) halo fraction times spin-independent  proton-neutralino scattering cross section, as a function of the mass, from Ref.~8 (color indicates standard relic densities).}
\vspace{-10pt}
\end{figure}

 Not only the relic density of WIMPs but their characteristic speed before structure formation in the Universe can differ in standard and non-standard pre-BBN cosmological models.	 
 If kinetic decoupling  (the moment when the exchange  of momentum between WIMPs and radiation ceases) happens during the reheating phase, WIMPs can have much smaller characteristic speeds, i.e.  be ``ultracold"~\cite{gg-UCWIMPS}, with free-streaming lengths several orders of magnitude smaller than in the standard scenario. Much smaller DM structures  could thus be formed, a fraction of which may persist as minihaloes within our galaxy and be detected in indirect DM searches. WIMPs may be much ``hotter"  than in standard cosmologies too, they may even be warm DM instead of cold. This could happens if WIMPs are produced with a large energy through late $\phi$-decays and subsequently  do not lose energy in interactions with the thermal bath~\cite{warm}.

 ``Visible" sterile neutrinos ($\nu_s$), i.e. those that could be found soon in neutrino experiments, would also be remnants of the pre-BBN era, if  their mass is $m_s > 10^{-3}$ eV (they are produced through oscillations mostly at $T\simeq 13 {\rm\,MeV} (m_s/ 1{\rm\,eV})^{1/3}$). In order to be found in experiments, these $\nu_s$ would necessarily have mixings with active neutrinos large enough to be overabundant, and thus be rejected, in standard cosmologies. In low $T_{RH}$ models the early Universe abundance of visible $\nu_s$  could be reduced enough for them to be cosmologically acceptable~\cite{visible-sterile}.
 
 Cosmological scenarios with a low $T_{RH}$ are more
complicated than the standard one. Although no consistent all-encompassing model
of this nature exists at present,  different aspects have been studied (e.g. for baryogenesis see Ref.~12)
and suggest  that a coherent scenario could
be produced if an experimental indication would lead us to it. 
Finding  a DM particle or a ``visible" sterile neutrino, whose
existence would contradict the standard assumptions about the pre-BBN
era, would give us not only invaluable information for particle
physics, but also an indication of enormous relevance in cosmology: it
would tell us that the standard assumptions must be modified. Low  reheating temperature models
provide an interesting alternative.

\section*{Acknowledgments}
This work was supported in part by DOE grant DE-FG03-91ER40662 Task C.


\section*{References}
\begin{thebibliography}{99}

  
\bibitem{regions} J.~R.~Ellis, K.~A.~Olive, Y.~Santoso and V.~C.~Spanos,
 {\it Phys.\ Lett.}\ B {\bf 565}, 176 (2003).
 
\bibitem{benchmarks} M.~Battaglia {\it et al.}
  {\it Eur.\ Phys.\ J.}\  C {\bf 33}, 273 (2004);
 B.~C.~Allanach {\it et al.},
  [arXiv:hep-ph/0202233];
  M.~Battaglia {\it et al.}
  [arXiv:hep-ex/0603010].


\bibitem{Moroi-etc} T. Moroi and L. Randall, {\it Nucl.\ Phys.}\ {\bf
B570}, 455 (2000); M. Fujii, K. Hamaguchi, {\it Phys.\ Rev.}\ D {\bf 66},
083501 (2002); D. H. Lyth, E.D. Stewart,
{\it Phys.\ Rev.}\ D {\bf 53}, 1784 (1996).

\bibitem{hannestad} 
  M. Kawasaki, K. Kohri, and N. Sugiyama, {\it Phys.\ Rev.\ Lett.}\ {\bf 82}, 4168 (1999); {\it Phys.\ Rev.}\ D {\bf 62}, 023506 (2000); S.~Hannestad, {\it Phys.\ Rev.}\ D {\bf 70}, 043506 (2004).
 
\bibitem{kinetion} P.~Salati,
  {\it Phys.\ Lett.}\ B {\bf 571}, 121 (2003):
  S.~Profumo and P.~Ullio,
 {\it  JCAP} {\bf 0311}, 006 (2003).
  
\bibitem{gg} G.~B.~Gelmini and P.~Gondolo,
 {\it  Phys.\ Rev.}\  D {\bf 74}, 023510 (2006).
  
\bibitem{ggsy} G.~Gelmini, P.~Gondolo, A.~Soldatenko and C.~E.~Yaguna,
  {\it Phys.\ Rev.}\  D {\bf 74}, 083514 (2006).
    
\bibitem{ggsy-2} G.~B.~Gelmini, P.~Gondolo, A.~Soldatenko and C.~Yaguna,
  {\it Phys.\ Rev.}\  D {\bf 76}, 015010 (2007).

\bibitem{gg-UCWIMPS} G.~B.~Gelmini and P.~Gondolo,
  arXiv:0803.2349 [astro-ph].

\bibitem{warm} W.~Lin et al.
  {\it Phys.\ Rev.\ Lett.}\  {\bf 86} (2001) 954:
  J.~Hisano, K.~Kohri and M.~Nojiri,
  {\it Phys.\ Lett.}\ B {\bf 505}, 169 (2001);
G.~Gelmini and C.~Yaguna,
  {\it Phys.\ Lett.}\  B {\bf 643}, 241 (2006).


\bibitem{visible-sterile}  G.~Gelmini, S.~Palomares-Ruiz and S.~Pascoli,
  {\it Phys.\ Rev.\ Lett.}\  {\bf 93}, 081302 (2004):
  G.~B.~Gelmini,
  {\it Int.\ J.\ Mod.\ Phys}.\  A {\bf 20}, 4670 (2005);
   G.~Gelmini, E.~Osoba, S.~Palomares-Ruiz and S.~Pascoli,
  arXiv:0803.2735 [astro-ph].
  
  \bibitem{Dolgov:2002vf}
  A.~D.~Dolgov, K.~Kohri, O.~Seto and J.~Yokoyama,
  {\it Phys.\ Rev.}\  D {\bf 67}, 103515 (2003).


\end{thebibliography}
\end{document}